\documentclass[english]{article}
\usepackage[latin9]{inputenc}

\makeatletter

\providecommand{\tabularnewline}{\\}

\makeatother

\usepackage{babel}
\begin{document}

\title{Expectation Value of $\sigma$(1).$\sigma$(2)-Wave Functions Don't
Matter }

\author{Larry Zamick\\
\\
 \textit{Department of Physics and Astronomy}, Rutgers University,
Piscataway, New Jersey 08854 }
\maketitle
\begin{abstract}
We consider the expectation value of the quantity {[}3+ $\sigma$(1).$\sigma$(2){]}/4
. This has a value +1 for 2 nucleons with spin S=1 and zero for S=0.
We show that for the jj coupling 2 particle configuration {[}j(1)
j(2){]}$^{J}$ the expectation value has the structure A+B J(J+1)
where A and B are constants. We then show that for a 2proton-2neutron
configuration with total angular momentum I the expectation value
per pair is independent of the details of the wave function and has
a similar structure A' +B' I(I+1) with B'=B/6.
\end{abstract}

\section{Introduction}

In this work we wish to study the spin contents of nuclear wave funcitons
in a single j shell model space.For convenience we use the spin one
projection operator {[}3+ $\sigma$(1).$\sigma$(2){]}/4 which has
a value of one for two particles coupled to S=1 and zero for 2 particles
coupled to S=0. However it does not make any difference what linear
combinaiotn of a constant and a term proportional to $\sigma$(1).$\sigma$(2)
one takes.

\section{Two particles in jj coupling}

We consider, in jj coupling, the two particle expectation values of
the spin one projection operator: P1(J)= \textless{} (jj)$^{J}$\textbar{}
{[}3+ $\sigma$(1).$\sigma$(2){]}/4\textbar{} (jj)$^{J}$\textgreater{}.
Here j is an abreviation for (l 1/2)j. As mentioned above this interaction
is non vanishing only for S=1. The total angular momentum is J.We
can use the unitary 9j coefficinets to evaluate this matrix elment.
The expression is

P1(J)= $\sum$ $_{L}|$\textless{}(l, 1/2)$^{j}$(l,1/2)$^{j}$\textbar{}
(ll)$^{L}$(1/2,1.2)$^{S=1}$\textgreater{}$^{J}$ \textbar{}$^{2}$

We here give the results for j= f$_{7/2}$.

The values P1(J) from J=0 to J=7 are shown in Table I.

.

.%
\begin{tabular}{|c|c|}
\hline 
J & Spin Eexpection\tabularnewline
\hline 
\hline 
0 & 0.42857\tabularnewline
\hline 
1 & 0.44898\tabularnewline
\hline 
2 & 0.48980\tabularnewline
\hline 
3 & 0.55101\tabularnewline
\hline 
4 & 0.63263\tabularnewline
\hline 
5 & 0.73469\tabularnewline
\hline 
6 & 0.85714\tabularnewline
\hline 
7 & 1.00000\tabularnewline
\hline 
\end{tabular}

.

Note that for J=7 we have L=6 and S=1,i.e. it is a pure S=1 state
so we are not surprised that P1(7)=1.

We can rewrite this as P1(J) = 0.42857 + 0.010204 J (J+1).

We can obtain these results in a simpler way. In a single j shell
we can replace $\sigma$ by g$_{j}$ j. For j=L+1/2 we have g$_{j}$=1/j
, whilst for j=L-1/2 g$_{j}$ = -1/(j+1). Thus, for j= L+1/2, $\sigma$(1).$\sigma$(2
is replaced by j(1).j(2)/j$^{2}$

Now j(1).j(2) is equal to {[}j(1)+j(2){]}$^{2}$/2 -j(j+1) = J(J+1)/2
-j(j+1). Putting this all together we have for j=l+1/2

P1(J)= 3/4-(j+1)/4j +J(J+1)/(8j$^{2}$).

\section{A system of 2 protons and 2 neutrons ($^{44}$Ti).}

We next consider the expectation value of P1(J) for a system of 2
protons and 2 neutrons i.e. $^{44}$ Ti.

The J=0$^{+}$wave functions for 2 protons and 2 neutrons in $^{44}$
Ti with the MBZE interaction {[}1{]} are given in the appendix. For
any four particle angular momentum I they are of the form

$\sum$ D$^{I}$(J$_{P}$ J$_{N}$) {[} (jj)$^{J_{P}}$(jj)$^{J_{N}}${]}$^{I}$ 

Here D(J$_{P}$ J$_{N}$) is the probability amplitude that the 2
protons couple to J$_{P}$ and the 2 neutrons to J$_{N}$. The normalization
condiiton is:

$\sum|D^{I}(J_{P}$J$_{N}$)\textbar{}$^{2}$=1

The expectation value of {[}3+ $\sigma$(i).$\sigma$(j){]}/4 for
J=0$^{+}$ states is given , per pair (there are 6 pairs) by

EXV= 1/6 (C1+C2)

C1= $\sum_{J_{P}}$ $_{J_{N}}$\textbar{}D$^{I}$(J$_{P}$ J$_{N}$)\textbar{}$^{2}$(
P1(J$_{P}$) + P1(J$_{N}$) )

C2=4.0{*}$\sum$$_{J_{_{A}}}$ F(J$_{A}$) P1(J$_{A}$)

where F(J$_{A}$) =$\sum$$_{J_{B}}$ \textbar{} $\sum_{J_{P}}$$_{J_{N}}$\textless{}(jj)$^{J_{P}}$(jj)$^{J_{N}}$\textbar{}
(jj)$^{J_{B}}$(jj)$^{J_{A}}$\textgreater{}$^{I}$ D$^{I}$(J$_{P}$
J$_{N}$)\textbar{}$^{2}$

We have what is an initially surpriing result. The expectation per
pair that S=1 is independent of the values of D$^{I}$(J$_{P}$ J$_{N}$
) . The value in fact is 0.643 for all 4 states. Thus the J=0$^{+}$
have more S=1 than S=0 but one cannot talk of correlations since the
results do not depend on the details of the wave functions.

It should be noted that although EXV does not depend on the wave function
components the quantities C1 and C2 do. This is shown in Table II.

.Table II The spin expectation for J=0$^{+}$ per pair .

.%
\begin{tabular}{|c|c|c|c|c|}
\hline 
 & State 1 & State 2 & State 3 (T=2) & State 4\tabularnewline
\hline 
\hline 
C1/6 & 0.488 & 0.452 & 0.428 & 0.400\tabularnewline
\hline 
C2/6 & 0.155 & 0.191 & 0.215 & 0.243\tabularnewline
\hline 
Total & 0.643 & 0.643 & 0.643 & 0.643\tabularnewline
\hline 
\end{tabular} 

.

We should compaere this with the value of S for a single nucleon.
For j=L+1/2 we have S=1. Thus the results for I=0$^{+}$ represent
some spin supression.

. We have repeated the calculations for states with higher angular
momentum . We find again that for any given the wave functions don't
matter. The expectation of P1(I) does depend on I.

\section{Simpler Considerations}

In the above we addressed the problem of the expectation value of
the spin operator. The essence of the problem , however, can be dealt
with more simply if we just take the expectation of J(J+1) whre J
is the 2 particle angular momentum e.g. 0 for J=0,1{*}2 for J=1,2{*}3
for J=2 e.t.c. It turns out that for any J=0$^{+}$ state in $^{44}$Ti
the non-normalized expectaion value is 126 , which is the same as
8j(j+1) (with j=7/2 in this case). The normalized value is 21. For
a 4 particle state of angular momentum I the expectation value is
simply 126+ I (I+1). 

We can see better what is happening if we treat J(J+1) as 2-body interaction
and perform a matrix diagonalizaton. The we find that all states of
a given I are degenerate and the spectrum in the 4 particle sysytem
is the same as for the 2 particle system ,namely E(I)= I(I+1) + constant.
The multi-degeneracy can easily be explained by the fact that the
2-body J.J interaction between basis states (J$_{P}$,J$_{N}$) and
(J$_{P}$' J$_{N}$') vanishes unless J$_{P}$= J$_{P}$' and J$_{N}$=
J$_{N}$'. In more detail:

The interaction summed over all pairs is 1/2 $\sum$j(t) . $\sum$j(q)
-1/2 $\sum$j(t)$^{2}$. The second term is a constant and can be
ignored as far as mixing is concerned.The first term , acting on a
4 particle state of total angular momentum I yields an eigenvalue
I(I+1). Clearly one will not get off diagonal matrix elements from
this term either. This leads to the multi-degenerate states of a given
angular momentum. The eigenvalues have the structure I(I+1) + constant.

If we diagonalize rather P1(J) and divide by the number of pairs (in
this case 6) we get the spin probability per pair. The maximum possible
value of this quantntity is one. We show the results in Table III. 

TableIII Pair Spin Probability for 2 protons and 2 neutrons in the
f$_{7/2}$ shell.

.%
\begin{tabular}{|c|c|}
\hline 
 & Pair Spin Prob.\tabularnewline
\hline 
\hline 
0 & 0.6428\tabularnewline
\hline 
1 & 0.6462\tabularnewline
\hline 
2 & 0.6530\tabularnewline
\hline 
3 & 0.6632\tabularnewline
\hline 
4 & 0.6768\tabularnewline
\hline 
5 & 0.6938\tabularnewline
\hline 
6 & 0.7141\tabularnewline
\hline 
7 & 0.7380\tabularnewline
\hline 
8 & 0.7652\tabularnewline
\hline 
9 & 0.7959\tabularnewline
\hline 
10 & 0.8446\tabularnewline
\hline 
11 & 0.8680\tabularnewline
\hline 
12 & 0.9081\tabularnewline
\hline 
\end{tabular}

. 

. Note that wheres with 2 nucleons the pair spin extectation value
is A+B J(J+1), whilst with 2 protons and 2 neutrons it is A' +B' I(I+1)
with B'=B/6. The values of A,A', B and B' are respectively 0.4286,
0.6428,0.0102 and 0.0034.

.

\section{Conclusions}

In talking about the amount of S=0 and S=1 content in nuclear wave
functions one must take care to separate results that do depend on
correlations and those that do not. One often hears the phrases''
S=0 J=0 pairing'' and ``S=1 T=0 pairing ``which imply correlation
dependent results. In the context in which these phrases are used
they may well be. justified But one should make sure that this is
the case. In this work we offer a conterpoint in whch the results
do not depend on the detailed wave functions.Our result 0.643,as the
pair probability of S=1 in any J=0+ in the f$_{7/2}$ shell model
space,is the same for all 4 J=0$^{+}$ states, 3 with isospin zero
and one with isospin 2,. That the pair spin probability is independent
of the wave functions in our single j sjell model space is true for
all I. Not surprisingly this quantitiy increases with incresing I
as 0.6428+ 0.0102 I(I+1).For I=12$^{+}$ we come close to the maximum
value of one. 

The value of this work is that it sets up a base for comparison. If
one wants to talk about S =1 correlations in a nucleus one shoud compare
the results with those here  shown in which the results do not depend
on thewave funcitons. The fact that the expectation of the spin projection
operator is here greater than 0.5 is simply due tot he fact that we
are in a j=L+1/2 model space and is not due to any subtle correlations.

\section{Appendix}

In TableIII we give the J=0$^{+}$wave functions in $^{44}$Ti with
the MBZE interaction {[}1{]}

.Table III Wave functions of J=0$^{+}$ states in $^{44}$Ti witht
he MBZE interaction.

.%
\begin{tabular}{|c|c|c|c|c|}
\hline 
Energy (MeV) & 0.00000 & 5.58610 & 8.28402$^{**}$ & 8.78750\tabularnewline
\hline 
\hline 
J$_{P}$ J$_{N}$ &  &  &  & \tabularnewline
\hline 
0 0 & 0.78776 & -0.35240 & -0.50000 & 0.07248\tabularnewline
\hline 
2 2 & 0.56165 & 0.73700 & 0.37268 & 0.04988\tabularnewline
\hline 
4 4 & 0.72080 & -0.37028 & 0.50000 & -0.75109\tabularnewline
\hline 
6 6 & 0.12340 & -0.44219 & 0.60092 & 0.65431\tabularnewline
\hline 
\end{tabular}

. The 3rd state has isospin T=2, all the others T=0.


\begin{thebibliography}{1}
\bibitem{key-1} A.Escuderos, L.Zamick and B.F. Bayman, arXiv:nucl-th/0506050.\end{thebibliography}
\end{document}